\documentclass[manuscript]{acmart}
\usepackage{enumitem}
\usepackage{graphicx}

\renewcommand\footnotetextcopyrightpermission[1]{} 

\def\participantnum{21 }
\AtBeginDocument{%
  }

\setcopyright{acmlicensed}
\copyrightyear{2018}
\acmYear{2018}
\acmDOI{XXXXXXX.XXXXXXX}
\renewcommand\footnotetextcopyrightpermission[1]{}
\makeatletter
\gdef\@copyrightpermission{}
\makeatother

\begin{document}

 \fancypagestyle{firstpagestyle}{%
    \fancyhf{}%
    \fancyhead[LE,LO]{\thepage}
    \fancyhead[RO,RE]{\shortauthors}
    \renewcommand{\headrulewidth}{0pt}%
    \renewcommand{\footrulewidth}{0pt}%
    \fancyfoot{}%
  }%
  \fancypagestyle{standardpagestyle}{%
    \fancyhf{}%
    \fancyhead[LE,LO]{\thepage}
    \fancyhead[RO,RE]{\shortauthors}
    \renewcommand{\headrulewidth}{0pt}%
    \renewcommand{\footrulewidth}{0pt}%
    \fancyfoot{}%
  }%
  \pagestyle{standardpagestyle}%

\title{VoiceMorph: How AI Voice Morphing Reveals the Boundaries of Auditory Self-Recognition}

\author{Kye Shimizu}
\email{kyeshmz@media.mit.edu}
\orcid{0000-0002-5017-6966}
\affiliation{%
  \institution{MIT Media Lab}
  \city{Cambridge}
  \country{USA}
}

\author{Minghan Gao}
\email{mg2328@cornell.edu}
\orcid{0000-0003-2302-2408}
\affiliation{%
  \institution{Cornell University}
  \city{Ithaca}
  \country{USA}}
\affiliation{%
  \institution{MIT Media Lab}
  \city{Cambridge}
  \country{USA}
}

\author{Ananya Ganesh}
\email{ag125@wellesley.edu}
\orcid{0009-0000-8817-1290}
\affiliation{%
  \institution{Wellesley College}
  \city{Cambridge}
  \country{USA}
}
\affiliation{%
  \institution{MIT Media Lab}
  \city{Cambridge}
  \country{USA}
}

\author{Pattie Maes}
\email{pattie@media.mit.edu}
\affiliation{
 \institution{MIT Media Lab}
  \city{Cambridge}
  \country{USA}
}

\renewcommand{\shortauthors}{Shimizu et al.}

\begin{abstract}
    This study investigated auditory self-recognition boundaries using AI voice morphing technology, examining when individuals cease recognizing their own voice. Through controlled morphing between participants' voices and demographically matched targets at 1\% increments using a mixed-methods design, we measured self-identification ratings and response times among 21 participants aged 18-64.

Results revealed a critical recognition threshold at 35.2\% morphing (95\% CI [31.4, 38.1]). Older participants tolerated significantly higher morphing levels before losing self-recognition ($\beta$ = 0.617, p = 0.048), suggesting age-related vulnerabilities. Greater acoustic embedding distances predicted slower decision-making ($r \approx 0.5-0.53, p < 0.05$), with the longest response times for cloned versions of participants' own voices.

Qualitative analysis revealed prosodic-based recognition strategies, universal voice manipulation discomfort, and awareness of applications spanning assistive technology to security risks. These findings establish foundational evidence for individual differences in voice morphing detection, with implications for AI ethics and vulnerable population protection as voice synthesis becomes accessible.

\end{abstract}

\begin{teaserfigure}
  \includegraphics[width=\textwidth]{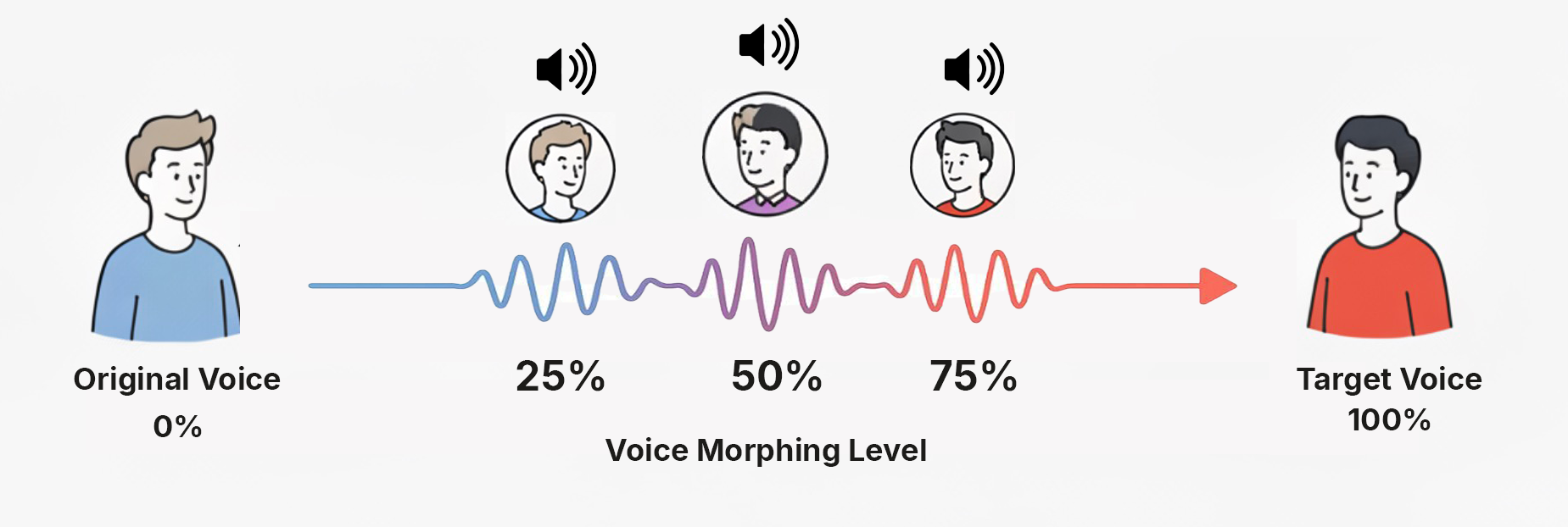}
  \caption{2 individuals with their voice identity being morphed together using neural voice morphing technology. }
  \Description{2 individuals with their voice identity being morphed together using neural voice morphing technology. Midway points show different levels of morphing.}
  \label{fig:teaser}
\end{teaserfigure}


\maketitle

\section{Introduction}
Before we can write, read, or even walk, we use our voices as a means of communication, and a way to interact with the world around us. This intimate relationship we have with our voice continues through our lives, from our first cry as newborns to conveying complex emotions. Through our voices, we project personality, emotional states, and countless other aspects of our own identity \cite{lavan_model_2023}. With the advancement of voice synthesizing technology, we can now create and manipulate vocal identities and the content being expressed. These emerging capabilities open new questions of how subtle modifications to our voice affect interpersonal dynamics, social cognition and perception, with broader implications ranging from accessibility in healthcare to risks of deception and misinformation in society \cite{wairagkar2025instantaneous}.


While prior research has shown that individuals struggle to identify whether a voice is artificially generated or not \cite{barrington_people_2025, muller_human_2022}, studies have mainly been focused on binary distinctions of authentic and manipulated voices. Little is known about the specific boundaries of auditory self-recognition when voices are gradually morphed.
Traditional methods could modify specific acoustic parameters, but only produced discrete changes to isolated voice characteristics. The advancement of deep learning-based voice cloning technology allows for neural voice synthesis that represents voices in high-dimensional latent spaces. Unlike digital signal processing (DSP) techniques, these models allow for smooth interpolation between speaker identities, creating a continuous spectrum of transformed voices. This presents a fundamentally different perceptual challenge: rather than simply detecting whether a voice is synthetic or not, individuals must now face voices that mix characteristics of themselves and others.

At the individual level, the perception of one’s own voice adds another layer of complexity, as the self-voice occupies a unique psychological status. As we process self-voice through both air and bone conduction pathways, this multimodal processing results in a perceptual mismatch when individuals hear recordings of their own voice \cite{orepic2023bone}. Research shows that the self-voice elicits distinctive neural responses \cite{graux2015my} and is prioritized in attention and recognition tasks \cite{kirk2025listen}. At the same time, listeners often show a self-enhancement bias, rating their own voices more positively in attractiveness \cite{hughes2013like, peng2020mechanism}. This special role of the self-voice in identity and self-perception highlights why its alteration through technologies like cloning may have unique cognitive and social consequences.

Extending from self-perception to social perception, prior research has shown that individuals tend to respond more favorably to voices that are acoustically similar to their own, suggesting a perceptual link between vocal similarity and social affiliation \cite{hughes2013like}. Recent studies have utilized voice cloning technologies to investigate how people perceive and socially evaluate synthetic voice replicas \cite{rosi2025perception, barrington_people_2025, 10.1145/3537674.3554742}. In parallel, past research in HCI has extensively explored voice-based systems and their usability in various contexts \cite{fang_leveraging_2025, dualvoice, liplearner, gong2001shall, 10.1145/286498.286815, cambre2020choice, Waterworth1985SyntheticCHI, TheEffectTaskSynthSpeechCHI2000}.

However, these lines of research largely treat voice identities as fixed categories of generated or authentic, rather than exploring the dynamic space that emerges when voices are able to morph between identities. Unlike fully synthetic voices that may trigger suspicion, morphed voices that blend familiar characteristics with unfamiliar elements could exploit trust mechanisms. Foundational work by Reeves and Nass demonstrates that humans automatically apply social rules to synthetic voices, treating them as real people \cite{mediaequation}, suggesting that morphed voices could exploit these unconscious social responses while remaining undetectable. This gap is critical as AI voice technologies become increasingly accessible and deployed across high-stakes contexts: AI-generated synthetic voices impersonating President Biden to suppress voter turnout in New Hampshire's 2024 primary \cite{seitz-wald_fake_2024}, emergency dispatch centers using AI assistants to handle 911 calls \cite{temkin_911_2025}, and corporate leaders envisioning "digital twins" that can attend meetings and make decisions on their behalf \cite{patel_zoom_2024}. Beyond these institutional risks, voice synthesis also enables new forms of personal identity manipulation and social engineering. Commercial platforms like ElevenLabs now offer real-time voice morphing capabilities that allow users to change gender, age, and accent of the target voice \cite{elevenlabs_reimagine_2025}. As AI systems move towards more personalized and adaptive interactions, it is important to understand the cognitive thresholds that govern when our sense of vocal self begins to blur.





To this end, we present a systematic investigation into how AI voice morphing reveals the boundaries of auditory self-recognition. Our approach employs controlled voice morphing between participants' authentic voices and demographically matched target voices to examine self-identification accuracy, response times, and subjective experiences. Through a mixed-methods design combining quantitative self-identification ratings with qualitative interviews, we reveal how individuals navigate the perceptual boundaries between self and other when encountering morphed versions of their own voice. 

Our findings contribute to understanding the cognitive mechanisms underlying auditory self-recognition and provide insights for the ethical design of voice synthesis technologies by revealing how alterations to self and familiar voices affect recognition and evaluation, underscoring the importance of authenticity in user trust and social perception.

\section{Related Work}

\subsection{Technical Foundations in voice identity manipulation}

Technology has transformed our ability to manipulate voice identity from statistical voice conversion methods to modern neural approaches. 
Vocoder toolkits such as STRAIGHT \cite{kawahara_straight_2006}, WORLD \cite{morise_world_2016}, STRAIGHTMORPH \cite{belin_straightmorph_2024}, DAVID \cite{rachman_david_2018},  PRAAT \cite{boersma2025praat} and others enable precise decomposition and manipulation of speech signals into components such as pitch and spectral envelope. These techniques have been used throughout HCI and cognitive science experiments such as to study voice attractiveness \cite{ferdenzi2013voice}, voice naturalness in emotional voice morphs \cite{nussbaum2023perceived}, how voices affect emotions \cite{aucouturier_covert_2016}, and self voice recognition through auditory feedback systems \cite{arakawa_digital_2021}. Such approaches have allowed researchers to systematically isolate and manipulate specific vocal characteristics and properties to understand their perceptual and cognitive effects.
However, previous research has shown that individuals adjust vocal properties such as pitch and speech rate in response to the social and situational context, including the nature of their relationships \cite{hazan2010does, zraick2006effect}. This is further reinforced in cognitive science theories such as the Communication Accommodation Theory \cite{communication-accomidation}, which suggests that speakers unconsciously modify vocal properties depending on the conversational partner. 
Modern neural approaches have fundamentally transformed voice identity manipulation by enabling control at a more holistic, high-dimensional level. By representing voices as latent vectors, this methodology enables the generation of interpolated voices between source and target speakers in the embedding space \cite{pani2023voice}. Previous research has highlighted a rapid transition in the field towards end-to-end models and speech foundation models such as GPT‑4o, capable of real-time expressive voice generation \cite{azzuni2025voice}. This shift towards neural approaches offers the opportunity to investigate continuous voice identity transformations and their
perceptual boundaries.

\subsection{Auditory Self-Recognition and Self-Voice Perception}

Understanding how individuals perceive and recognize their own voice has been well documented through cognitive theoretical foundations. The Forward Model Theory \cite{wolpert_computational_nodate, tourville_diva_2011} illustrates that when we speak, the brain predicts how our voice will sound. If the sound is different from the prediction, the brain will detect the error and adjust accordingly. This process helps explain why self-voice recognition is fundamentally different from other-voice recognition, as the brain treats self-produced speech in a distinct way \cite{behroozmand_error-dependent_2011}. Yet, prior research has shown that we often misjudge the spatial origin of our own voice \cite{wen_over-estimation_2022}, as well as that people are generally less accurate in recognizing their own voice compared to others \cite{candini_who_2014}. The Predictive Coding Theory \cite{friston_predictive_2009} conceptualizes the brain as continuously comparing predictions to incoming sensory information. As a result, we find a reduced level of activity in the superior temporal gyrus area due to successful prediction matching \cite{andics_neural_2010}. Realistic voice clones may intrude these established prediction patterns, which can produce a novel neural response that blurs self-other boundaries, as it is unable to successfully predict whether the voice is self-produced. Finally, the Social Identity Theory \cite{turner_social_1979} illustrates how the voice serves as a fundamental marker for social categorization, with temporal analyses revealing that physical characteristics such as age and gender are processed early (\~120ms), while social and trait-related cues emerge later (\~260ms) \cite{lavan_time_2024}.



\subsection{Interpersonal self-other boundaries}

Research increasingly recognizes that cognition is not confined to the brain but emerges from embodied engagement with the environment. Studies have shown that altering both visual and auditory cues of one’s own body can significantly change self-identification, highlighting how sensory manipulation blurs self–other boundaries and reshapes self-perception \cite{tajadura2017embodiment, fiedler2023embodiment, tajadura2012person}. 

Building on this, voice is often described as the most embodied instrument of human expression. Its production is rooted in bodily processes, yet its perception is socially and technologically mediated. Recent work in vocal performance has shown that identity is co-constructed through the interplay of body, technology, and perception \cite{baumann2023embodied}. Technological processing such as voice morphing or cloning can detach voice identity from the body or reassign it, altering how individuals relate to their own sound \cite{pani2023voice, deutschmann2011gender}. 

In the context of technological interventions in voice perception, empirical studies show that users’ reactions to voice morphing are strongly dependent on their pre-existing relationship with their voice \cite{johnson2021expectancy}. For example, research on avatar-mediated voice morphing found that users who like their own voice prefer their unaltered voice, while those dissatisfied with their voice are more likely to favor morphed voices that approach an idealized self-representation \cite{okano2022avatar}. This highlights how morphing technologies can mediate identity negotiation, enabling individuals to explore alternative vocal selves.

While identity negotiation is often framed individually, voice morphing also influences how others perceive and categorize voices, reshaping social recognition and evaluation. By altering acoustic cues, voice morphing modifies two key dimensions of perception: familiarity, or how known and easily recognized a voice feels, and self-relevance, or the degree to which a voice is tied to one’s sense of self, reshaping how voices are recognized and evaluated \cite{fontaine2017familiarity, stevenage2024familiarity, pinheiro2023attention}. Listeners show higher recognition accuracy for familiar voices and process them with less interference \cite{kanber2022highly, holmes2020speech}, with advantages often tied to close relationships \cite{domingo2020benefit}. Yet even unfamiliar voices can trigger self-prioritization under certain conditions \cite{kirk2025listen}. Self-relevance, in turn, modulates these judgments: people typically find their own voices more attractive \cite{hughes2013like}. At the same time, technological alterations introduce new dynamics where listeners acquainted with the speaker (self or friend) prefer the authenticity of original voices, whereas unfamiliar listeners often favor polished qualities of the cloned voices\cite{rosi2025perception}. Human-like qualities further boost likability \cite{kuhne2020human}.

\section{Problem Statement and Research Questions}
Building upon the theoretical foundations, we aim to investigate how morphing affects auditory self recognition.

\subsection{Research Questions}\label{RQs}

\begin{enumerate}[label=\textbf{RQ\arabic*:}, leftmargin=*, widest=RQ4:]
    \item \textbf{Voice Morphing Threshold and Latent Distance Relationship}
    \begin{itemize}
        \item \textit{Problem:} At what morphing percentage do participants fail to recognize morphed versions of their own voice, and how does this threshold relate to latent vector distance in the embedding space?
        \item \textit{Hypothesis:} We predict a sharp perceptual "cliff" where self-identification ratings drop significantly, with this threshold correlating with latent vector distance metrics in the neural embedding space.
        \item \textit{Method:} Participants rate self-identification (1-7 Likert scale) across 1-100\% morphing levels. We measure correlation between identification ratings and Euclidean distances of latent vectors (Section \ref{procedure}).
    \end{itemize}
    
    \item \textbf{Personality Modulation of Perceptual Boundaries}\label{RQ1}
    \begin{itemize}
        \item \textit{Problem:} Which Big Five personality factors modulate individual differences in voice self-recognition thresholds?
        \item \textit{Hypothesis:} High Openness will extend tolerance for morphed voices (resulting in a higher threshold), while high Neuroticism will reduce tolerance (resulting in a lower threshold), based on cognitive flexibility and sensitivity to familiar self-voice characteristics.
        \item \textit{Method:} Mini-IPIP\cite{donnellan_mini-ipip_2006} personality assessment correlated with individual morphing thresholds and self-identification rating patterns (Section \ref{quantitative}).
    \end{itemize}
    
    \item \textbf{Response Time Patterns Across Morphing Levels}\label{RQ2}
    \begin{itemize}
        \item \textit{Problem:} How do response times change as voices transition across the morphing spectrum from self to other?
        \item \textit{Hypothesis:} Based on Predictive Coding Theory\cite{friston_predictive_2009}, absolute conditions (0\% and 100\% morphing) will trigger faster responses than ambiguous mid-range morphed voices requiring extended cognitive processing.
        \item \textit{Method:} Response time measurement for self-identification judgments across all morphing conditions, with analysis of RT patterns relative to morphing percentage (Section \ref{quantitative}).
    \end{itemize}
    
    \item \textbf{Subjective Experience of Voice Morphing}\label{RQ4}
    \begin{itemize}
        \item \textit{Problem:} What subjective experiences and reasoning strategies do participants report when encountering morphed versions of their own voice?
        \item \textit{Method:} Semi-structured interviews following classification task, with thematic analysis of participant descriptions, uncanny valley effects, and decision-making processes (Section \ref{qualitative}).
    \end{itemize}
\end{enumerate}

\section{Methods}
\subsection{Power Analysis}

To determine appropriate sample sizes for our voice morphing study, we conducted power analysis using effect size recommendations from previous research specific to speech research \cite{gaeta2020examination}, small (d = 0.25, r = 0.25), medium (d = 0.55, r = 0.40), and large (d = 0.95, r = 0.65) effects. We estimated medium-large effects based on the strong perceptual manipulation involving clear self-other voice boundaries. Our final sample of N = 21 participants provides 67\% power to detect medium effects in linear mixed-effects models (RQ1, RQ3), 48\% power for medium effects in regression analyses of perceptual thresholds, and 44\% power for medium correlations between embedding distances and behavioral outcomes (RQ1), with >90\% power for large effects across all analyses.



\subsection{Experiment}

We conducted this research in-person rather than online for several reasons. Firstly, we wanted to ascertain that the quality of the voice sample obtained from the participant was high, ensuring no multitasking or background interference, and creating a control environment with the recording and playback equipment (Section \ref{equipment}). Moreover, the semi-structured interview format required dynamic interaction and contextual follow-up questions that are more feasible in person than through remote questionnaires.

\subsection{Procedure}\label{procedure}

\begin{figure}[ht!]
\centering
\includegraphics[width=0.9\textwidth]{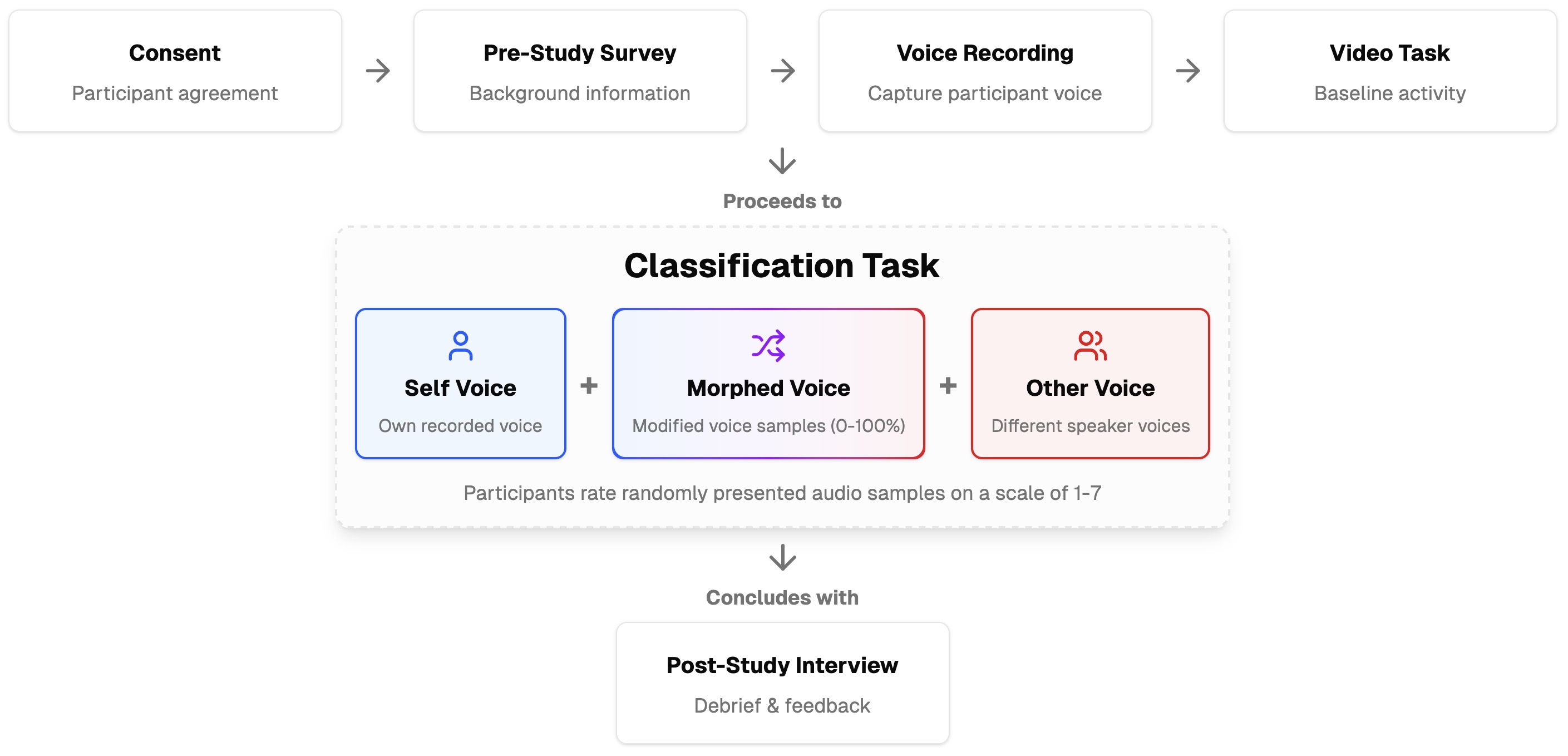}
\caption{Overview of experiment procedure \label{overflow}}
\Description{A flow chart that shows the timeline of the experiment. 1. Each participant is asked for their consent in the experiment. 2. Each participnat fills out a pre-study survey that asks about their background information as well as demographic information. 3. Each participant's voice is recorded. 4. A baseline activity of a video task is introduced. 5. Participants proceed to a classification task where they are asked to classify on a series of audio files, where they rate from 1-7 on whether the voice sounds like them. The audio stimuli is a corpus consisting of their own voice, morphed voices, and other voices. 6. Lastly, participants take a post study interview about the experiment.}
\end{figure}

\begin{enumerate}\phantomsection\label{methods}
  \item \textbf{Consent} Participants who were eligible through the  screening process (Section \ref{participants}) were told about the experiment and guided through the consent form. Participants signed the consent form before participating in the experiment. We note that in the consent form, there is no aforementioned use of "cloning" or other words to describe the neural synthesis, but rather use "manipulated" when referencing synthetically generated voices to not introduce bias before the experiment. 
  \item \textbf{Pre Study Survey} Before the start of the study, participants were asked to complete the Mini IPIP, a shorter form of the Big Five Questionnaire (Section \ref{miniipip}) \cite{donnellan_mini-ipip_2006}.
  \item \textbf{Voice Recording} At the start of the session, each participant was asked to speak a few simple sentences from the Rainbow Passage \cite{rainbow-sentences} into a microphone connected to a laptop. These recordings were subsequently used to create different versions of their voice, including a computer-generated (cloned) version using Cartesia  (Section   \ref{cartesia}).
  
  \item \textbf{Video Task} To reduce potential short-term memory effects from participants recalling the acoustic features of their own recorded voice, participants were instructed to watch a video including speech in between the recording and experiment. By introducing a video task filled with speech, we reduce the likelihood that participants rely on immediate auditory recall or echoic memory. This interval ensures participants cannot
use fresh memory of their vocal features as a reference point when
making self-voice judgments. The filler video content featured both male and female speakers to accommodate participants across the gender spectrum and minimize potential gender-based voice priming effects \cite{blades_gable_ted_2017}. 
  \item \textbf{Classification Task} Participants wore a pair of closed ear adjustable headphones, while listening to a series of audio clips, including both the original recordings of themselves and the modified voices based on that. For each clip, participants were asked to decide whether the voice sounded like their own or not and to indicate their answer on a tablet or computer (Section \ref{classfication}).
  \item \textbf{Exit Interview} After completing the task, participants engaged in a brief semi-structured interview (Section \ref{qualitative}). The questions, adapted from prior work \cite{itsnotarepresentation}, focused on participants’ impressions of the system, appropriate contexts for use, and potential improvements. During this session, the overall experiment paradigm was also revealed.
\end{enumerate}

\subsubsection{Classification Task}\label{classfication}

Each participant classification task sequence was a randomized sequence of 16 participant unmodified recorded voice excerpts of the Rainbow Sentence spoken in their own voice, 16 "other" voice excerpts, and a sequence of morphed voices that were interpolated from 1\% to 100\% in 1\% increments, totaling 131 different samples. Trial order was fully randomized for each participant using the random module in Python with no constraints on sequence. This randomized order was implemented to ensure that any time-related changes in participant performance such as decreased attention are distributed across experiment conditions and other order effects. 
The baseline trials serve three primary functions: establishing individual recognition accuracy benchmarks for self and other voices, providing anchors for interpreting morphed voice ratings, and validating that participants can reliably distinguish between self and other voices. 

We calculated the minimum number of baseline trials (self and other voice conditions) needed with the standard error of mean formula: $SE_{\bar{x}} = \sigma/\sqrt{n}$, where $\sigma$ represents the expected standard deviation of recognition rating within each baseline condition and $n$ is the number of trials. Based on pilot data, we estimated $\sigma \approx 1.2$ for both self and other voice conditions on our 7-point Likert scale. To achieve a standard error of $\leq 0.3$ scale points (ensuring 95\% confidence intervals of $\pm 0.6$ points), the required sample size is $n = (\sigma/SE)^2 = (1.2/0.3)^2 = 16$ trials per condition.


During the task, a fixation of a white cross was shown for 2 seconds, followed by the audio stimuli. The audio stimuli length was based on the audio file used; original (mean=10.17 s, SD=0.05), cloned m=8.40 s, SD=0.55), or morphed m=7.89s, SD=1.03). The participants are asked to respond to the similarity of the voice using the 7-point Likert scale within 5 seconds. The recording of the response time occurs after this response screen is shown, and a confirmation of what the participant had selected is shown shortly afterward. A blank gap screen is shown in between each of the trials for 2 seconds. The design of the trials are structured so that the "echoic memory" \cite{cognitive-psychology} of the last trial will not affect the following trial.

\begin{figure}[H]
\centering
\includegraphics[width=0.95\textwidth]{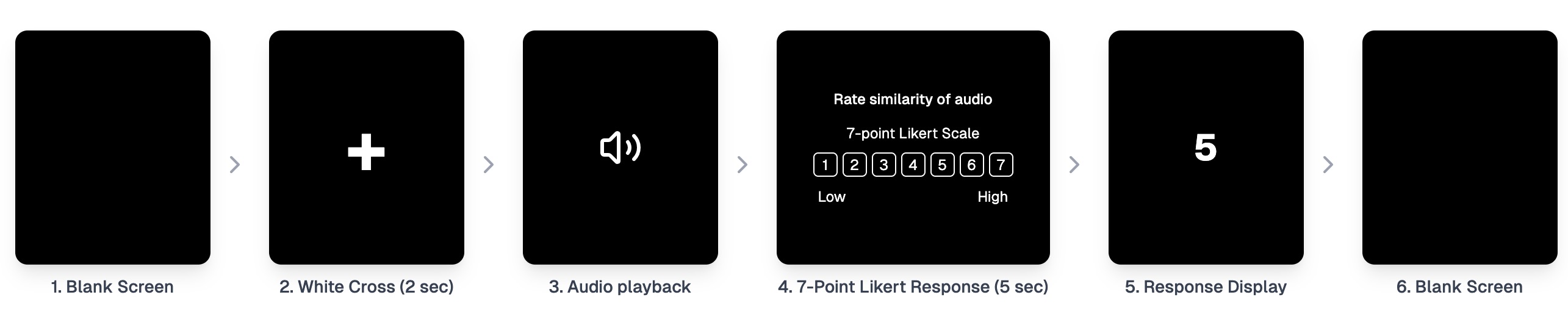}
\caption{Overview of classification task. Participants repeated this sequence for all 131 samples. \label{overflow}}
\Description{For the classification task, each participant is presented a series of screens: 1. A blank black screen is shown, 2. A white cross appears on the screen for 2 seconds, 3. Audio is played back and randomly chosen from the audio stimuli 4. Participants are asked to rate the audio with a 7 point likert response scale 5. There is a screen that shows which number the participant has chosen 6. There is another blank screen to add a gap between audio stimuli. This sequence repeats until all audio stimuli are presented}
\end{figure}

Each participant is given 3 practice trials to familiarize themselves with the scoring system, adjust the headphones, and practice answering the questions. After that, a reminder message is shown on the screen to indicate the beginning of the main trials.

\subsubsection{Clone Apparatus}\label{cartesia}

The Cartesia SDK \cite{noauthor_cartesia_nodate} was selected for its ability to generate high-quality voice clones with embedding data attached, perform rapid voice cloning with minimal data, interpolate between voice identities in latent space, and stream synthesized audio in real time. These features enabled us to efficiently create and manipulate personalized voice clones, generate intermediate voices between participants and target identities, and deliver generated audio in low latency. More specifically, the API version {2024--11--13} and {sonic--2--2025--03--07} model were used, as future versions of the SDK employ a different architecture model that does not allow for mixing or embedding generation.

Open source models such as Tortoise TTS \cite{betker_better_2023} offer more interoperability within the voice model, but at the cost of cloning and data training time. Other proprietary models such as Elevenlabs \cite{elevenlabs-no-date} allow for high fidelity instant voice cloning and have been featured in previous HCI research \cite{fang_leveraging_2025, brade_speakeasy_2025, sew_user_2025, ada-2024}, but do not allow for the usage of mixing or acquiring embedding data from the voice model.

Original recordings averaged 10.17s (SD = 0.05), while synthesized outputs were consistently shorter: cloned voices (M = 8.40s, SD = 0.55) and morphed voices (M = 7.89s, SD = 1.03). Importantly, no correlation existed between individual participants' input and output durations, indicating that the synthesis process applied consistent transformations regardless of original recording length. This duration reduction likely reflects the neural network's optimization of prosodic features, adjusting the speech rate, pausing patterns, and syllable timing rather than proportional scaling of the input audio.

\subsubsection{Other Voice Generation}
The other voice pairs were chosen from the Common Voice Dataset \cite{ardila_common_2020} which was selected over alternative datasets for its comprehensive speaker metadata, including speaker age, gender, and accent information. We adhered to V1.1 Metadata schema, and filtered from the voices using the criteria below. 

Demographic matching (Section \ref{participants}) followed the V1.1 Metadata schema of the Common Voice Dataset, controlling for self recognition judgments by pairing each participant with other voices of the same gender, same age range, US English accents, and high voice quality (UTMOS-v2 scores >4.0). This matching protocol ensures that morphing effects reflect self-other boundaries rather than artifacts from demographic voice differences. Previous research has shown that age and gender are processed within 120ms of voice onset \cite{lavan_time_2024}, making demographic matching critical for isolating self-recognition mechanisms from basic demographic categorization.

We limited the selection to speakers with a U.S. English accent in order to align with our recruitment population, which consisted primarily of university-affiliated participants from the United States. Additionally, we excluded the other accent labels in the dataset due to two key concerns: (1) the practical challenge of recruiting participants across all represented accent groups, and (2) the difficulty in verifying the authenticity and consistency of accent labels in the dataset. This approach allowed us to maintain linguistic consistency and demographic alignment between the dataset and the participant population.

To ensure that only high-quality audio samples were used from the Common Voice dataset, we applied the UTMOSv2 system \cite{baba_t05_2024} to automate the estimation of the Mean Opinion Score (MOS) \cite{ITU2017P10} metric for each clip. While UTMOSv2 was optimized for synthetic speech evaluation, it was trained on datasets containing both synthetic and natural speech samples, making it capable of evaluating speech quality across domains. While we acknowledge potential domain transfer effects from synthetic to natural speech that UTMOSv2 is trained upon, we note that underestimation of natural speech would yield stricter filtering criteria.

We retained only those samples with predicted scores above 4, which is aligned from previous research filtering scores above 4 as high quality speech \cite{zhou_mos-fad_2024, ITU2017P10}. This threshold excludes samples with potential acoustic artifacts that could confound voice morphing perception while acknowledging the inherent limitations of MOS-based quality assessment (Section \ref{mos}).

To make sure that the participant's responses were reflective of genuine perceptual differences in voice characteristics rather than playback volume inconsistencies of the audio files, all audio stimuli were normalized using FFMPEG \cite{ffmpeg2025}. The loudness of each file was calculated according to the EBU R128 standard and adjusted using the loud norm filter with parameters, target loudness of \textminus18 LUFS, a maximum true peak of \textminus1 dBTP, and a loudness range of 11 LU.

\subsubsection{Apparatus} \label{equipment}

Audio data was captured using a Shure SM7B dynamic microphone in a studio environment, with participants wearing Audio Technica ATH-M20x closed-back headphones to provide passive noise isolation and minimize  environmental acoustic interference. The microphone was interfaced through a Focusrite Scarlett 18i20 USB audio interface. All audio signals were digitally sampled at 48 kHz with 16 bit.

\subsubsection{Generating morphing sequence}
Voice morphing was performed by linearly interpolating between 192-dimensional voice embeddings using the Cartesia Voice Mix API endpoint \cite{cartesia_mix_api}. Following Cartesia's documented linear interpolation approach \cite{cartesia_embeddings}, we applied the formula:
\begin{equation}
C = (1-\alpha)A + \alpha B
\end{equation}
where $A$ and $B$ represent the 192-dimensional embedding vectors of the original cloned voice (from) and the destination cloned voice (to), respectively, and $\alpha$ serves as the interpolation coefficient. When $\alpha = 0$, the result equals $A$ (original voice), and when $\alpha = 1$, the result equals $B$ (target voice). The morphing parameter $\alpha$ was systematically varied from 0 to 1 in increments of 0.01, generating 101 distinct voice samples per participant.

\subsubsection{Prompt}

For the voice recording section of the experiment, we used the Rainbow Passage \cite{rainbow-sentences} over alternative standardized speech corpora, such as the Harvard Sentences \cite{noauthor_ieee_nodate}, for its phonetically balanced structure, neutral content, content length, and content consistency between sentences that allowed participants to speak without utterances in several sentences. As our technical pipeline required approximately 10 seconds of continuous speech per participant for voice cloning, two sentences were sequentially read from the passage, allowing for natural sounding speech. 

\subsubsection{Conditions}

The experimental design employed three distinct auditory conditions to systematically examine voice perception and self-recognition processes:

\begin{enumerate}
    \item \textbf{Original}: Participants' unmodified recorded voices served as the baseline stimulus, establishing a control for authentic self-voice perception.

    \item \textbf{Original-and-Cloned Mix}: Participants were presented with morphed vocal stimuli created through computational voice synthesis, blending their authentic voice with that of another individual to generate a hybrid vocal identity. This condition was specifically designed to examine the perceptual and cognitive effects of ambiguous vocal ownership and graduated self-similarity.

    \item \textbf{Other Voice}: Demographically matched (gender and age) but acoustically distinct non-self voices were employed, providing a comparative framework for investigating the boundaries of voice recognition and self-identification processes.
\end{enumerate}

We included the original recordings as controls to establish baseline self-recognition under naturalistic conditions. While this introduces a quality difference from synthesized voices, it reflects real-world scenarios where individuals compare potentially manipulated voices against their natural voice. Future work could isolate morphing effects by using purely synthesized baselines.

\subsection{Measures}

\subsubsection{Participant Information}\label{participants}

We recruited \participantnum participants from a university through message groups, social networks, mailing lists, and word of mouth. All participants were U.S. residents and first completed a screening questionnaire on Qualtrics to assess demographic information and medical history. Eligibility criteria included being over 18 years old, having no past or current affiliation with the research group, reporting English as a native language, and having no speech or auditory impairments that may hinder task performance. Individuals with a known diagnosis of any mental health disorder, neurological condition, or learning disability were excluded. Participants received 25 USD in compensation and the study was approved by the University Institutional Review Board (IRB).


Out of the \participantnum participants, 6 identified as men and 15 identified as women. Participants' ages ranged from 18 to 64 (mean=29.0, SD=13.9). All participants were self-reportedly fluent in English, with 11 being fluent in one additional language and 2 in two additional languages. These languages included Spanish, Mandarin, Hindi, French, Uzbek, Italian, Tamil, Vietnamese, and Japanese. We acknowledge some of the limitations of the participant population and report upon limitations in Section \ref{sample-size}.

\subsubsection{Acoustic Similarity}
To quantify similarity between audio clips, pairwise Euclidean distances of the latent vectors generated by Cartesia were computed directly on the unnormalized representations of each recording. For each participant, \textit{within-speaker} distances were calculated as all pairwise comparisons among that participant’s own clips (including source, morphed, and cloned versions), and \textit{between-speaker} distances were calculated as comparisons between a participant’s clips and those of all other participants. The \textit{delta} metric was defined as the difference between the mean between-speaker and mean within-speaker distances.

\begin{figure}[H]
\centering
\includegraphics[width=0.7\textwidth]{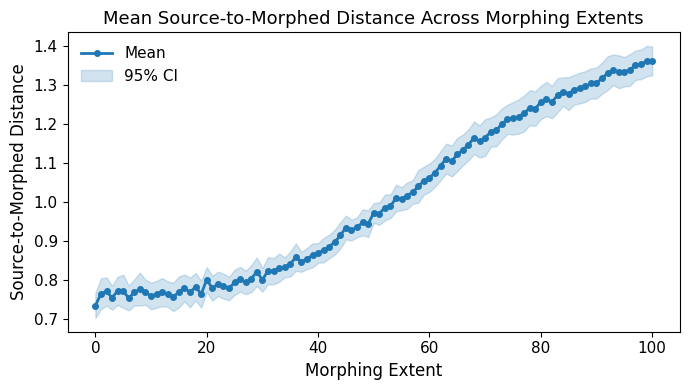}
\Description{Figure 4 shows how the mean source-to-morphed distance changes across morphing extents. At the beginning of the continuum (0–20\%), distances remain relatively stable, hovering around 0.75–0.8, indicating that small morphing adjustments result in minimal deviation from the original source voice. From roughly 40\% onward, however, the distance begins to rise steadily, showing an almost linear increase as morphing progresses. By the time the morph reaches 100\%, the distance surpasses 1.3, reflecting substantial acoustic divergence from the source. The shaded band indicates the 95\% confidence interval, which stays relatively narrow across the continuum, suggesting a consistent pattern across participants.
Note that the relatively flat section at the lower morphing levels in Figure 4 reflects the fact that early morphing steps introduce subtle acoustic changes that are less perceptually distinct. In contrast, the sharper rise at higher extents signals that later morphing increments contribute disproportionately more to acoustic distance, magnifying the perceived departure from the source voice.
}
\caption{Mean source-to-morphed distance as a function of morphing extent (95\% CI) 
\label{overflow}}
\end{figure}

When examining distances between morphed voices and their original source, the trajectory increased with morphing extent, but not linearly. Instead, the curve followed a sigmoidal shape, with slower changes at the early and late morphing stages and a steeper rise near the middle of the continuum. This pattern indicates that embedding-based deviations reflect nonlinear perceptual changes as morphing progresses.

\begin{table}[H]
\centering
\caption{Descriptive statistics for within-speaker, between-speaker, and delta (between -- within) distances across participants ($n=21$).}
\Description{Table 1 summarizes descriptive statistics for within-speaker, between-speaker, and delta (between – within) distances across participants. On average, within-speaker distances were smaller (M = 0.758) and more tightly clustered, while between-speaker distances were larger (M = 1.220) with greater spread. The resulting delta measure averaged 0.462, reflecting a consistent gap between within- and between-speaker variability. A paired t-test confirmed this difference to be highly significant, t(20) = 24.10, p < .001, with a very large effect size (d = 5.26). These results establish clear separation in the embedding space between individual voice variability and cross-speaker differences.}
\label{tab:within-between-delta}
\begin{tabular}{l@{\hskip 6pt}cccccc}
\toprule
\textbf{Measure} & \textbf{M} & \textbf{SD} & \textbf{Med.} & \textbf{IQR} &
\multicolumn{2}{c}{\textbf{95\% CI}} \\
\cmidrule(lr){6-7}
 & & & & & \textbf{Low} & \textbf{High} \\
\midrule
Within-speaker  & 0.758 & 0.066 & 0.758 & 0.070 & 0.728 & 0.788 \\
Between-speaker & 1.220 & 0.086 & 1.203 & 0.140 & 1.181 & 1.259 \\
Delta (B -- W)  & 0.462 & 0.088 & 0.443 & 0.114 & 0.422 & 0.502 \\
\midrule
\multicolumn{7}{l}{\textbf{Test Statistics}} \\
Paired $t$-test & \multicolumn{6}{l}{$t(20) = 24.10, \; p = 3.0 \times 10^{-16}, \; d = 5.26$} \\
\bottomrule
\end{tabular}
\end{table}

\textbf{Across all $n = 21$ participants, within-speaker distances were substantially smaller than between-speaker distances} (Table~\ref{tab:within-between-delta}). On average, within-speaker clips clustered tightly ($M = 0.76$), while between-speaker clips were much farther apart ($M = 1.22$). The mean delta ($M = 0.46$) reflects this consistent separation, indicating that cross-speaker distances exceeded same-speaker distances by nearly half a unit in the embedding space. A paired-samples $t$-test confirmed the robustness of this effect, $t(20) = 24.10, \; p < .001$, with a very large effect size ($d = 5.26$). These results demonstrate strong speaker separation, with recordings from the same individual forming coherent clusters relative to cross-speaker comparisons. This pattern also supports the validity of the Cartesia embedding space: if clips from the same speaker failed to cluster, it would undermine the reliability of the embeddings for capturing voice similarity.

\subsubsection{Quality Assessment}

To characterize stimulus quality and guard against quality-driven confounds, clip-level Mean Opinion Scores (MOS; 1–5) were estimated with UTMOSv2 from raw waveforms for all stimuli (Original, Morphed, Cloned). Scores were merged with behavioral trials by filename and participant (participant ID extracted after \texttt{ID\_} in the MOS metadata). MOS was used only for descriptive and robustness checks; it was not included as a predictor in hypothesis-testing models.

Across all clips ($N=2163$), predicted quality was mid-range on average: mean MOS $= 2.64$ ($\mathrm{SD}=0.54$, 95\% CI [2.62, 2.66]), spanning 0.81–3.86 with the middle 50\% between 2.34 and 3.03. Most clips therefore clustered around the center of the scale, indicating consistent but not uniformly high quality. When summarized by participant, MOS values varied meaningfully. Means ranged from 1.99 to 3.09, with most participants between 2.5 and 3.0. This spread likely reflects a combination of recording conditions, speaker characteristics, and synthesis performance.

By morphing extent, MOS increased from approximately 2.08 at 0\% to above 3.0 at 100\%, with tight confidence intervals across levels, indicating fewer artifacts and higher predicted naturalness at greater morphing. To verify that MOS did not confound behavioral inferences, relationships with perceived own-voice similarity (1–7) and response time were visualized with LOWESS (frac $=0.4$) and summarized by correlation. Observed trends were shallow, with correspondingly small correlations (similarity: Pearson $r=-0.10$, $p=9.6\times10^{-9}$; response time: $r=-0.12$, $p=1.9\times10^{-10}$; RTs trimmed to 200–6000\,ms), accounting for roughly 1–1.5\% of variance. These checks indicate that audio quality is unlikely to drive the main morphing-level effects on perceived similarity or latency.

\begin{figure}[H]
\centering
\small
\includegraphics[width=0.8\textwidth]{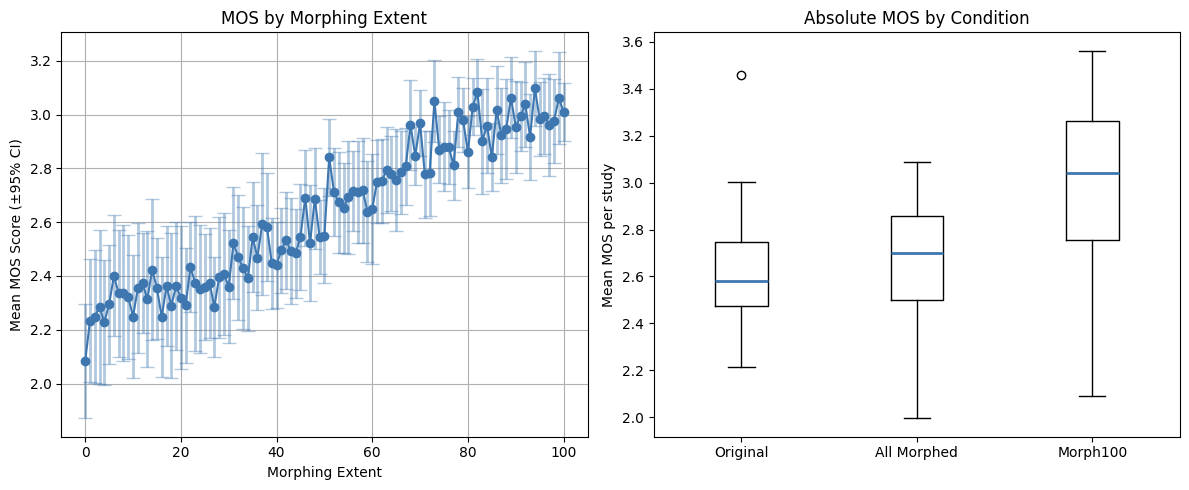}
\caption{MOS by morphing extent and condition. Left: condition means with 95\% CIs. Right: boxplots by condition.}
\Description{Figure 5 illustrates mean opinion scores (MOS) across morphing extents and conditions. On the left, MOS increases gradually with morphing extent, starting around 2.1 at 0\% and rising to approximately 3.0 by 100\%. Although the confidence intervals widen slightly, the overall upward trajectory indicates that more heavily morphed voices were rated, the higher the audio qualities are according to the MOS scale. On the right, boxplots summarize MOS distributions across conditions. The original voices show lower medians and tighter clustering, while fully morphed (100\%) voices have the highest median ratings but also greater spread, reflecting increased variability in judgments. The “All Morphed” condition falls between these extremes.}
\label{fig:mos-by-morph}
\end{figure}

\section{Results}

\subsection{Quantitative Analysis}\label{quantitative}

To analyze the participants' self-identification responses, we employed a quantitative analysis using a linear mixed effects model. The dependent variable is the self-identification rating, measured on a 7-point Likert scale. The model will include both categorical and continuous independent variables. Categorical variables will consist of gender, age group, and voice type (Self, Other, Morphed), with morphed voices matched to participants in gender and approximate age. Continuous predictors will include Mini-IPIP  scores, response time,  the degree of morphing (0\%-100\%), standardized Euclidean distance between the participant’s voice and the morphing target, and standardized demographic distance (age and gender) from the morphing target.

\subsubsection{Voice Boundary Thresholds and Age Effects - Addressing RQ1}

To address RQ1(\ref{RQs}), we employed a two-stage analytical
approach.
First, we applied LOWESS (Locally Weighted Scatterplot Smoothing) with smoothing fraction 0.4 to extract individual perceptual thresholds where self-recognition ratings crossed
below 4.0 on our 7-point scale. LOWESS was chosen over parametric approaches (logistic regression, psychometric functions) as participants exhibited highly variable response patterns, and also does not force artificial mathematical constraints on the responses. 
Secondly, we used OLS regression via Python's statsmodels \cite{seabold2010statsmodels} to examine demographic predictors of the extracted thresholds (lowess\_T).

This two-stage analysis provides initial evidence for significant age-related effects in voice boundary perception: \textbf{older participants demonstrated higher perceptual thresholds} ($\beta$ = 0.617, p = 0.048), before they stopped recognizing their own voice. Gender effects were non-significant (p = 0.346), though the overall model explained 25\% of variance in threshold locations (R² = 0.252).

\begin{table}[H]
\Description{This table presents OLS regression results examining predictors of voice boundary thresholds,
defined as the morphing percentage where self-recognition drops below 4.0. The analysis reveals
that age is a significant positive predictor (β = 0.617, p = 0.048), indicating that older
participants maintain self-recognition at higher levels of voice morphing compared to younger
individuals. Gender shows no significant effect (β = -9.563, p = 0.346), suggesting male and
female participants do not differ substantially in their voice boundary thresholds. The model
explains 25.2\% of the variance in voice recognition thresholds (R² = 0.252) based on 17
observations, though the overall model significance is marginal (F = 2.357, p = 0.131). These
findings suggest that aging may enhance voice self-recognition robustness, possibly due to more
stable voice identity representations or reduced sensitivity to acoustic distortions, while
gender appears to play a minimal role in determining voice boundary perception thresholds.
}
\centering
\small
\caption{OLS Regression Results: Voice Boundary Thresholds (lowess\_T)}
\label{tab:threshold-regression}
\begin{tabular}{lcccccc}
\toprule
\textbf{Variable} & \textbf{Coeff.} & \textbf{Std. Err.} & \textbf{t} & \textbf{P>|t|} &
\multicolumn{2}{c}{\textbf{95\% CI}} \\
\cmidrule(lr){6-7}
 & & & & & \textbf{Lower} & \textbf{Upper} \\
\midrule
Intercept & 30.202 & 8.802 & 3.431 & 0.004** & 11.324 & 49.080 \\
Gender  & -9.563 & 9.809 & -0.975 & 0.346 & -30.602 & 11.477 \\
Age & 0.617 & 0.285 & 2.165 & 0.048* & 0.006 & 1.228 \\
\midrule
\multicolumn{7}{l}{\textbf{Model Statistics}} \\
R-squared & 0.252 & & Adj. R-squared & 0.145 & & \\
F-statistic & 2.357 & & Prob (F-stat) & 0.131 & & \\
Observations & 17 & & AIC & 144.9 & & \\
\bottomrule
\end{tabular}


\smallskip
\textit{Note.} * p < 0.05, ** p < 0.01. Gender coded as: Female = 0, Male = 1. Dependent variable: voice boundary threshold (morphing \% where self-recognition < 4.0).

\vspace*{-0.4\baselineskip}
\end{table}

\subsubsection{Effect of Latent Distances on Thresholds and Response Time - Addressing RQ1}

To examine the relationship between acoustic embedding measures and behavioral outcomes, correlations were conducted between within-speaker distances, between-speaker distances, and their difference (delta) against both perceptual thresholds and response times. No reliable associations emerged between any of the embedding metrics and perceptual thresholds: within-speaker similarity showed essentially no correlation (\(r \approx 0.02\), n.s.), while both between-speaker distances (\(r \approx 0.19\), n.s.) and the delta metric (\(r \approx 0.17\), n.s.) were weak and non-significant. In contrast, response times displayed systematic dependencies on embedding distances. \textbf{Participants with greater within-speaker distances required significantly longer to respond (\(r \approx 0.5\), \(p < .05\)), and the same was true for between-speaker distances (\(r \approx 0.53\), \(p < .05\))}, with the latter showing a slightly stronger effect. The delta metric, however, did not predict response time (\(r \approx 0.15\), n.s.). These findings indicate that while embedding-based similarity does not determine where participants cease to recognize voices as their own, it does influence the speed of decision-making.

\begin{figure}[H]
\centering
\includegraphics[width=0.95\textwidth]{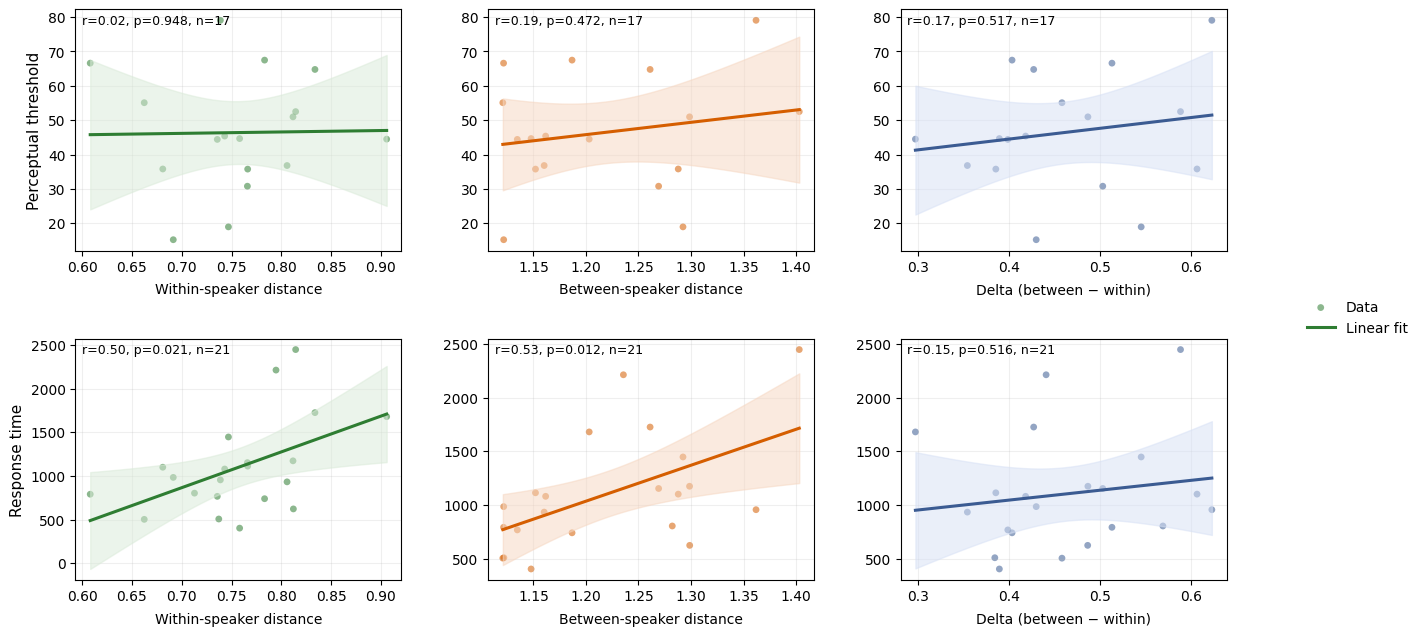}
\caption{Correlations between embedding distances and behavioral outcomes (perceptual threshold and response time)
\Description{Figure 6 shows correlations between embedding distances and behavioral outcomes. The top row plots perceptual thresholds against within-speaker, between-speaker, and delta distances. Across all three, no significant relationships were observed, as slopes remained flat and correlation values were small. By contrast, the bottom row shows response times against the same embedding measures. Here, both within-speaker distance (r = .50, p = .021) and between-speaker distance (r = .53, p = .012) exhibit significant positive correlations: participants with larger distances took longer to respond. The delta measure did not show a meaningful association with response time.}
\label{overflow}}
\end{figure}

Principal component analysis provided further insight into the structure of these embedding measures. The first two principal components accounted for nearly all the variance (PC1 = 57.2\%, PC2 = 42.8\%), with between-speaker distances and the delta metric loading heavily on PC1, while within-speaker distances aligned strongly with PC2. This orthogonal separation confirms that within-speaker variability and cross-speaker contrast capture distinct, non-overlapping dimensions of the acoustic embedding space. Scatter plot analyses reinforced this pattern: regression lines for both within- and between-speaker distances showed positive slopes with response times, with a steeper gradient for between-speaker distances, indicating that greater cross-speaker separation tends to slow participant judgments more than internal variability. The delta metric again showed little explanatory power in either statistical or visual terms.

In general, the analyses suggest that \textbf{perceptual boundaries for self-voice recognition are not directly governed by raw embedding distances, but that larger distances, particularly between speakers, contribute to slower decision processes, potentially reflecting additional cognitive efforts and longer decision latencies}. The PCA results further underscore the need to separate within-speaker variability from cross-speaker contrasts, suggesting that recognition judgments emerge from an interplay between subtle, speaker-specific fluctuations and broader inter-speaker differences in voice quality.

\begin{table}[H]
\centering
\caption{Correlation Results: Embedding Measures and Behavioral Outcomes}
\Description{Figure 5 illustrates mean opinion scores (MOS) across morphing extents and conditions. On the left, MOS increases gradually with morphing extent, starting around 2.1 at 0\% and rising to approximately 3.0 by 100\%. Although the confidence intervals widen slightly, the overall upward trajectory indicates that more heavily morphed voices were rated, the higher the audio qualities are according to the MOS scale. On the right, boxplots summarize MOS distributions across conditions. The original voices show lower medians and tighter clustering, while fully morphed (100\%) voices have the highest median ratings but also greater spread, reflecting increased variability in judgments. The “All Morphed” condition falls between these extremes.}
\label{tab:correlation-results}
\Description{Table 3 presents correlations between embedding measures and behavioral outcomes. Perceptual thresholds did not correlate significantly with any embedding metric, indicating that boundary placement was not systematically tied to acoustic variability. By contrast, response times showed robust associations with both within-speaker (r = .49, p = .024) and between-speaker distances (r = .54, p = .012). Participants with greater distances required longer times to make recognition decisions, consistent with the idea that acoustic dissimilarity imposes additional cognitive effort. Delta measures showed no reliable links with either threshold or response time. Together, these findings suggest that while categorical thresholds remain stable, processing speed is sensitive to the degree of acoustic dispersion in the embedding space.}
\begin{tabular}{lcccccc}
\toprule
\small
\textbf{X Variable} & \textbf{Y Variable} & \textbf{n} & \textit{r} & \textbf{$\rho$} & \textit{$\rho$} & \textbf{p} \\
\midrule
Within-mean   & Threshold (lowess\_T) & 17 & 0.02  & 0.94  & 0.08  & 0.76 \\
Within-mean   & Response time         & 21 & 0.49* & 0.024 & 0.49* & 0.026 \\
Between-mean  & Threshold (lowess\_T) & 17 & 0.19  & 0.47  & 0.09  & 0.74 \\
Between-mean  & Response time         & 21 & 0.54* & 0.012 & 0.54* & 0.012 \\
Delta         & Threshold (lowess\_T) & 17 & 0.17  & 0.52  & 0.18  & 0.50 \\
Delta         & Response time         & 21 & 0.15  & 0.50  & 0.22  & 0.33 \\
\bottomrule
\end{tabular}

\smallskip
\textit{Note.} * p < .05 (two-tailed). Threshold = morphing percentage where self-recognition rating drops below 4.0.

\end{table}

\subsubsection{Personality Modulation of Perceptual Boundaries - Addressing RQ2}

To investigate whether Big Five personality traits modulate individual differences in self-voice recognition, we examined correlations between all five personality dimensions and both perceptual thresholds and categorization times. Contrary to our hypothesis, regression analyses revealed no significant relationships between personality traits and perceptual thresholds, despite response times showing considerable individual variation. Similar null results were obtained for LOWESS-derived thresholds (F(5,11) = 0.294, p = 0.907). The overall model predicting threshold boundaries was non-significant (F(5,11) = 0.294, p = 0.907, R² = 0.118).

\subsubsection{Effects of Morphing Level on Self-rating and Response Time - Addressing RQ3}

LOWESS trends (\textit{frac} = 0.4) show a monotonic decrease in perceived own-voice similarity (1--7) with increasing morphing. The overall LOWESS threshold, the morphing percentage at which the fit crosses a rating of 4, occurs at \textbf{35.2\%} (95\% CI \textbf{[31.4, 38.1]}; Fig.~8, top-left). Correlation analyses align with this pattern (Pearson \(r=-0.57\), \(p=2.49\times10^{-184}\), \(n=2{,}118\)). Response time exhibits a modest downward trend with morphing (Fig.~8, bottom-left), yielding a weak but reliable association (Pearson \(r=-0.09\), \(p=5.37\times10^{-5}\), \(n=2{,}019\)). 

The gradual sigmoid-shaped recognition curves observed in our data, rather than sharp cliff-like transitions, align with signal detection theory predictions for noisy perceptual judgments, where the perceptual signal-to-noise ratio gradually improves rather than creating discrete categorical boundaries.

Together, these trends clarify the perceptual dynamics of voice morphing. To probe these dynamics further, we examined effects under four experimental conditions. The original corresponded to the participants’ unaltered voice recordings. Morphed 0 denoted a cloned voice generated with a morphing level of 0, positioned at the closest point to the original voice on the morphing continuum between self and other. Morphed 100 represented a cloned voice fully transformed (100\%) into another speaker’s voice matched on demographic traits, and presented repeatedly throughout the session. Finally, Cloned also represented a fully transformed (100\%) other voice, but was presented only once, thereby functioning as a single-exposure counterpart to Morphed 100. 

Based on the four conditions, the results demonstrate that the perceived similarity is highest for Original (mean \(=6.74\), 95\% CI \([6.64, 6.84]\)), intermediate for Morphed~0 (\(4.36\) \([3.71, 5.01]\)), and at the low end of the scale for Morphed~100 (\(1.48\) \([1.03, 1.92]\)) and Cloned (\(1.63\) \([1.52, 1.74]\)). The per participant contrasts (BH-FDR) show large, reliable differences between Original and Cloned (\(d=6.36\), \(q<.001\)), between Original and Morphed~0 (\(d=1.95\), \(q<.001\)), and between Morphed~0 and Morphed~100 (\(d=2.36\), \(q<.001\)), whereas no significant difference emerged between Morphed~100 and Cloned (\(q=.52\)). For response time, Morphed~0 is descriptively slowest (mean \(=1{,}237\)~ms, 95\% CI \([752, 1{,}722]\)), while Original (\(847\)~ms \([774, 920]\)), Morphed~100 (\(946\)~ms \([575, 1{,}317]\)), and Cloned (\(926\)~ms \([837, 1{,}016]\)) are lower. No pairwise RT differences remained significant after FDR (all \(q\ge .15\)).

\begin{figure}[H]
\centering
\includegraphics[width=0.8\textwidth]{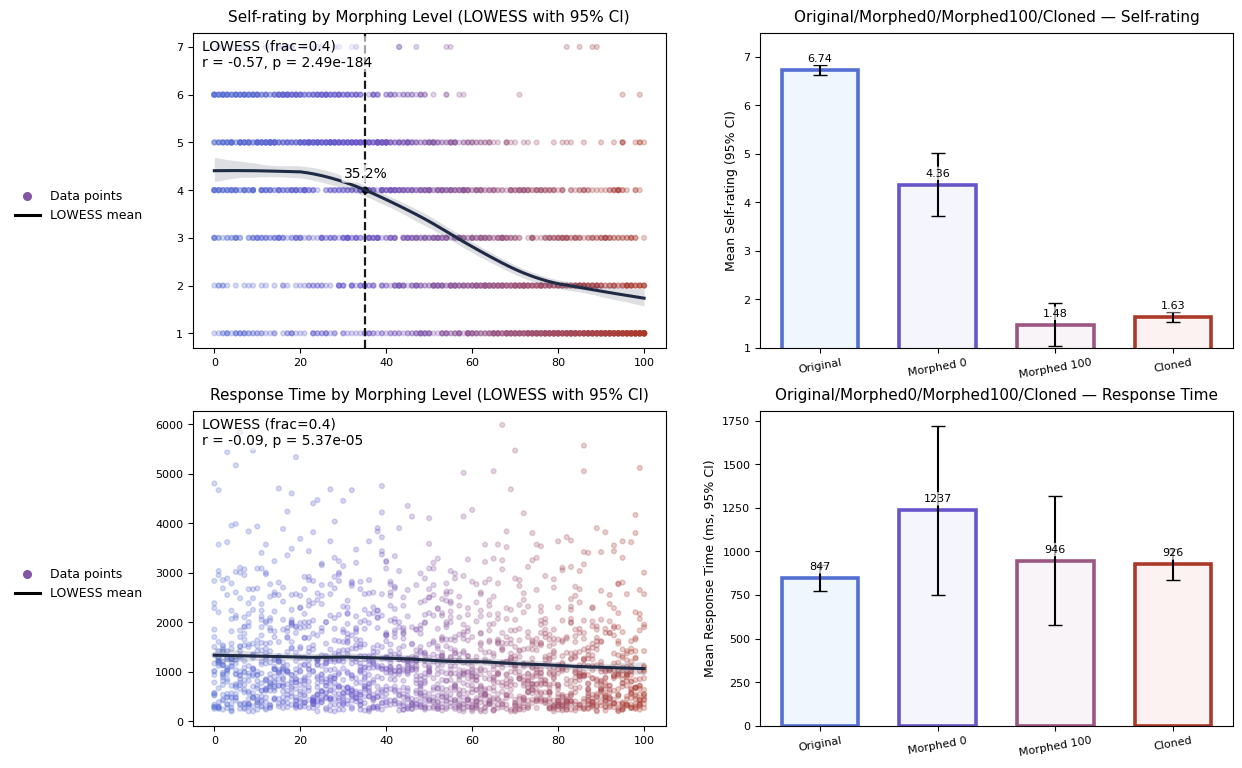}
\Description{Figure 7 shows self-ratings and response times as a function of morphing extent and condition. On the left, the LOWESS curve for self-ratings declines gradually with increasing morphing: participants rated voices as most self-like at 0\% morphing, with ratings dropping steadily toward 100\% morphing. The vertical line marks the group-level perceptual boundary at 35.2\%, where ratings cross the midpoint of the scale. In contrast, response times show little systematic variation across the morphing continuum, with a slight downward slope that is not strongly predictive.
The bar plots on the right highlight condition-level differences. Original voices received the highest self-ratings (M = 6.74), while both fully morphed (M = 1.48) and cloned voices (M = 1.63) were rated as least self-like. Morphed0 (the cloned baseline with 0\% morphing) occupied an intermediate position (M = 4.86). For response times, the slowest decisions occurred in the Morphed0 condition (M = 1237 ms), whereas original and cloned voices elicited quicker judgments (M = 847 ms and M = 926 ms, respectively). This pattern suggests that ambiguous stimuli near the self–other boundary demand more cognitive effort, while clearly self-like or other-like voices are classified more efficiently.
}
\caption{Self-rating and response time by morphing extent
\label{overflow}}
\end{figure}

\subsection{Qualitative Analysis}\label{qualitative}



In addition to the quantitative analysis, we conducted qualitative interviews with each participant through a semi-structured interview. The interviews were transcribed across all sessions. Thematic coding was done on the participants' comments collected during the post-experiment interview process. Approximately 78 open codes were generated from the transcripts and organized into 5 core themes, and iterated with input from other members of the research team. These domains encompass: \textbf{(1) Voice Identity \& Self-Concept, (2) Perceptual Processing \& Recognition, (3) Responses to Voice Manipulation, (4) Emotional \& Affective Dimensions, and (5) Technology \& Social Implications}. The interviews were assessed to provide further insight into participants’ subjective experiences, interpretations of the voice stimuli, and reasoning behind their self-identification ratings.

\subsubsection{Subjective Experience and Reasoning Strategies - Addressing RQ4}

Multiple participants expressed discomfort and awkwardness when hearing their recorded voice, revealing various dimensions of auditory self-recognition challenges. One participant captured the universal nature of this experience, stating: "\textit{I think nobody really likes to hear their voice.}" (P5), "\textit{I think my voice is high-pitched, and then the voice in my head sounds deeper}"  (P4). This is aligned with previous research that has examined an individual's mixed self perception of their own voice \cite{CHONG20241249.e19, Holzman1966TheVA}, as well as how a recorded voice sounds different from their own voice due to differences in bone conduction and eardrum vibration caused by bone conduction \cite{tonndorf1968}. 

Participants reported feeling unsettled by partial transformations pertaining to accents, as one participant described, "\textit{Like some of them, I feel like I had a British accent.}" (P1, P2, P3), which aligns with limitations with our dataset (Section \ref{limitations-dataset}).  Some participants also noted that the synthetic voices appeared to speak faster (P2, P6, P9) and that there was less 'stretching' (P10) of certain syllables, which made it easier to identify them. This observation could be attributed to the shorter average duration of the morphed voices, as described in \ref{methods}, resulting in a perceptibly higher speech rate.

Self-consciousness emerged as a response to the dissonance between self and morphed voices, as one participant expressed, "\textit{I was a little bit self-conscious at times because I thought that's not how I really sound}" (P5). The effect of this is compounded by uncanny valley effects (Section \ref{uncanny}). 

Participants voiced different implications and possible applications that the technology can inform across domains.  As participant P8, who works with ALS patients, explained: "\textit{Patients often lose their voice. And one of their goals is to get a synthesizer that actually recreates their own voice artificially. And they're getting better and better at doing it}". One participant (P7) noted how this could be used for anonymization or in security where only the individual could identify the authentic voice. Other participants were concerned about how such voice data could be easily harvested as well as dark patterns that may arise from the usage of mixed synthetic voices such as marketing (P4), blackmailing (P2), and scam (P4). 

\subsubsection{Prosodic Recognition Cues and Acoustic Validation - Addressing RQ4}

A participant, who was an English teacher for international students, outlined the difference of "\textit{ups and downs}" (P2) that the morphed voice differed from what they intended. This sentiment was echoed by multiple participants, who emphasized the importance of intonation as a defining characteristic of their voice. One participant noted, "\textit{I think different people have different intonations, so I can easily tell this is my, like, intonation or this is how I talk.}" (P2) and another observed that the morphed voices sounded like "\textit{another voice, possibly me, but with different intonation}" (P6). Others spoke about "\textit{the way they would say something}" (P4), implying that they used word emphasis and lexical stress as cues to rate similarity, with one participant noting that "\textit{There's some words I emphasize and some words that they emphasize on...and when I hear it, I instantly know that it's not me}" (P10). 

To assess whether these claims were quantifiable, we utilized Parselmouth \cite{jadoul2018parselmouth}, a Python interface for Praat \cite{boersma2025praat}, a widely adopted tool in computational phonetics and acoustic analysis. We extracted the frequency spectrum from each morphed audio stimulus and calculated the coefficient of variation (CV) of pitch by dividing the standard deviation of pitch by its mean. An ordinary least squares (OLS) linear regression revealed a small but statistically significant positive effect of the degree of morphing on pitch variability ($\beta$ = 0.0376, 95\% CI [0.0116, 0.0637], p = 0.0047). However, the explanatory power of the model was very limited, as indicated by the low coefficient of determination (R² = 0.003).

The results support the participants' reports, which highlighted the perception of a unique intonation in their own voice as a distinguishing factor. Although participants described this as something they believed to be an intrinsic and highly personal quality, the data indicate that these perceptions could correspond to measurable changes in pitch variation. This suggests that the subjective sense of self-voice is potentially grounded in specific and quantifiable acoustic features that can be altered to inhibit recognition. 

These insights contextualize our quantitative findings, showing that perceptual boundaries reflect both acoustic similarity and individual identity attachment.

\section{Discussion}


\subsection{Societal Implications}\label{social}

Synthetic voice technology has become integral to modern life, powering beneficial applications from voice assistants like Siri \cite{apple_siri_2011} and Alexa \cite{amazon_alexa_2014} to preserving vocal identity for ALS patients \cite{regondi_artificial_2025}. However, the same technological capabilities that drive these beneficial uses have also enabled severe real-world consequences for corporate fraud \cite{damiani_voice_deepfake_2019}, financial authentication bypasses \cite{cox_vice_bank_2023}, systematic bias against minority speakers \cite{koenecke_racial_2021}, and targeted scam operations \cite{ftc_ai_deception_2023, aivoiceclonedecieve}. 

While these documented incidents involve fully synthetic voices, neural voice morphing technology presents a largely unexplored area of research. Previous research has demonstrated that people show measurable preference and increased trust for voices that are acoustically similar to their own \cite{jaggy_ai-determined_2025}, as well as prioritization over other voices \cite{payne_perceptual_2021}. Related research on voice interfaces, such as GPS navigation systems, provides concrete evidence of how synthetic voice characteristics significantly affect user perception \cite{zhang_artificial_2025}, cognitive load \cite{listenbefore}, and decision-making \cite{LARGE201469e1}. 

Our findings on voice morphing thresholds contribute to understanding how self similarity bias can be leveraged without detection, raising awareness of manipulation in voice-based systems. These findings alongside previous research in the field highlight how subtle voice characteristics can unconsciously influence judgments about competence and trustworthiness. Understanding how voice morphing plays a role becomes critical in developing appropriate safeguards and frameworks to protect users while preserving the beneficial aspects of voice technology. 


\subsection{Cognitive and Perceptual Implications}
Our analysis shows that greater latent distances, including both within and between participants, are correlated with longer response times. This suggests that acoustic dissimilarity imposes additional cognitive load, reducing the ability of participants to categorize a voice as 'self' or 'other'. Importantly, this effect emerged not only when comparing across speakers but also within an individual’s own voice space. Participants with greater variability within the speaker required more time to decide, reflecting the increased difficulty in resolving self-recognition under conditions of acoustic divergence. This result is consistent with previous studies demonstrating that voice familiarity and self-similarity reduce cognitive efforts and lead to greater confidence during recognition \cite{mckenzie2021listen, plante2021processing}.

These findings illustrate that self-other boundaries are not only categorical but also temporally sensitive, with recognition latencies serving as behavioral markers of boundary negotiation. From a design perspective, it implies that while voice morphing and cloning technologies enable identity exploration and alternative self-representations, excessive deviation from one's baseline voice can impose processing costs, slow recognition and potentially undermine usability.

\subsection{Implications for HCI Design}\label{elder-hci}

Building on these perceptual dynamics, we consider how they translate into practical challenges and opportunities for system design. Cognitive load and recognition latency broadly shape usability, but these effects become especially salient for populations already facing age-related perceptual decline. Our preliminary findings extend this line of work by identifying age-related threshold differences. While recent elder voice research emphasizes personalization and familiarity \cite{ValdiviaL25, TorresEtAl19}, our identification of age-related threshold differences (Section \ref{quantitative}) and individual vulnerability windows suggests that current personalization approaches may inadvertently create security risks for elderly populations utilizing voice interfaces. 
Furthermore, as our finding indicates that participants respond faster to other voices compared to cloned versions of their own voice, voice interfaces that incorporate self-voice elements may inherently impose greater cognitive load on users \cite{ROSA2008204, HUGHES20101124}. This cognitive overhead may directly impact usability of systems that utilize modified versions of self voice for interventions \cite{fang_leveraging_2025, jo2024neural}. This is especially problematic for elderly users, who already face age-related cognitive decline \cite{impact-aging} and reduced ability to accurately assess their own auditory perception \cite{rogers_frequent_2012}. 

Recent work by Nautsch et al. (2021) demonstrates that speaker verification systems can be personalized based on individual acoustic features \cite{POHLHAUSEN2026101823}, while Byeon et al. (2022) show that users desire contextual voice customization ranging from improved versions of their own voice to completely different voices for identity protection \cite{VoiceThatSuits}. Building on this approach, our morphing threshold data could inform adaptive systems utilizing self-voice threshold. 


\subsection{Limitations and Future Work}

\subsubsection{Other Voice Dataset}\label{limitations-dataset}
While the Common Voice dataset provides diversity in demographics as well as comprehensive speaker metadata, there are practical constraints to the dataset, such as the audio equipment not being standardized across contributors and metadata information being self-reported. While our systematic demographic matching approach ensures methodological consistency, it represents a practical compromise between similarity control and feasibility rather than true perceptual equivalence.

\subsubsection{MOS Prediction}\label{mos}

While MOS provides a standardized quality metric, there are inherent biases to the usage of MOS and MOS predictors \cite{ragano_comparison_2023, streijl_mean_2016}. MOS scores are relative to the specific test context rather than an absolute measure of quality, which introduces range equalization bias \cite{cooper_investigating_2023}. Furthermore, our quality based filtering may reduce speaker diversity, as automated MOS predictors might systematically rate certain demographic groups higher, introducing demographic bias. Despite these limitations, the MOS threshold provides a practical threshold to systematically process thousands of voice audio samples in large scale datasets, while maintaining a standard for voice quality assessments.

\subsubsection{Sample Size \& Generalization}\label{sample-size}

The study's sample of \participantnum participants was recruited primarily from a university population, resulting in a group that was predominantly young and female. Although we believe that this sample size is sufficient for an initial investigation, replicating the study with a larger and more diverse participant pool will be valuable to further strengthen the generalization of our findings.

Furthermore, the study's restriction to English-speaking participants limits the extent to which these results can be applied in the context of speakers of other languages and cultural backgrounds, where intonation and other prosodic features may differ considerably. Future research should consider cross-linguistic investigations to address these limitations and further validate the observed effects.

\subsubsection{Model Limitations}

This current research leverages the instant cloning API that Cartesia implements in their SDK, allowing for cloning to be done using a short sample and without a long training step, with a trade-off that it might not be as accurate. Future work may consider the utilization of other models such as the open source Tortoise TTS \cite{betker_better_2023} and other models \cite{christop_cloneval_2025} to generate more realistic voice that further improve the understanding and replicability of the results. We were also not able to accommodate different types of accents that arise in the English Language because of the difficulties assessing the accuracy of the accents, but accommodating different types of accents may increase the familiarity of the person to how certain words are being said. 

\subsubsection{Novelty Effects}\label{uncanny}
This study presents a novel paradigm where participants listen to morphed versions of their own voice rather than purely synthetic speakers. While recent research has looked into the uncanny valley effect of synthetic voices, revealing complex dynamics, such as how we may perceive synthesized voices as more likeable \cite{kuhne2020human}, or lessened uncanny valley effects when participants are aware that the voice is synthetic \cite{fang_leveraging_2025}, our approach introduces a fundamentally different perceptual challenge. As our methodology introduces an experience where participants encounter gradual transformations of their own voice, there may be limitations to the extent that our findings reflect stable perceptual boundaries versus temporary responses to this new paradigm. Future longitudinal studies examining how these perceptions change with repeated exposure would help entangle these novelty effects.



\subsubsection{Experiment Design}
There are clear limitations of the current work as it relies on the participants reading aloud, which is known to elicit a different part of the brain that is associated, than in a setting where a participant is talking to another person \cite{kuhlen_brains_2017, andics_neural_2010}.
Further experiments in this field should accommodate how the recording of the voice is taken and what the context is because of this difference. Furthermore, this research did not ask about the participant's familiarity with these voice cloning technologies to prevent any bias before the experiment itself, but there are potential bias issues when a participant already has cloned their voice before and has gone through a similar process.

Our audio-only approach also limits comparison with research on identity morphing in other modalities. Recent work on using neural models for facial morphing \cite{shimizu_morphing_2023, human-latent} demonstrates similar threshold effects in visual identity recognition. Future work can incorporate multimodal approaches to understand how auditory and visual identity boundaries interact in real-world scenarios.

Furthermore, this work was conducted exclusively in English with participants who declared English as their native language. Future work could investigate perceptual differences across different languages, explore voice morphing perception for other individuals (comparing familiar with unfamiliar voices), and explore how conversational agents might extend this research into broader interpersonal contexts.

\subsubsection{Prompt Selection}
Our prompts also came from the Rainbow Sentence dataset, which is designed to be a phonetically balanced dataset, but future work could look into how different emotional prompts or voices may change this boundary of self across different emotions.

\section{Conclusion}

This exploratory study investigated auditory self-recognition boundaries by morphing participants' voices with demographically matched targets while measuring self-identification ratings, response times, and personality traits. By utilizing voice cloning technology to morph two voices together, the study revealed a critical vulnerability in human auditory self-recognition that has profound implications for our AI-mediated future. By systematically mapping the boundaries of voice morphing detection, the results show that the overall recognition threshold at which individuals no longer recognize the morphed voice as their own is around 35.2\% of the morphing level. Among participants from different age groups, older adults tolerate higher levels of morphing before losing self-recognition ($\beta$ =0.617, p = 0.048). Through further examining the pairwise Euclidean distance of the latent vectors we acquired from voice embeddings, we concluded that larger acoustic embedding distances led to slower decision-making processes. In terms of personality, our analysis contradicted the original hypothesis and showed no predictive power for recognition thresholds. Across morphing levels, recognition was slowest for the 0\% morph (the cloned version of participants’ own voice), indicating that voices highly similar to one’s own impose greater cognitive load during recognition. Qualitative interviews highlighted universal voice discomfort, concerns about emotional mismatch in morphed voices, and participant awareness of potential applications ranging from ALS voice preservation to fraud vulnerabilities.

While our university-based sample (N=21) limits generalizability, these results establish foundational evidence for age-related differences in voice morphing detection that warrant further investigation. Future research should examine whether these perceptual boundaries can be shifted through training, explore cross-cultural and linguistic variations, and investigate similar threshold phenomena in other modalities of self-recognition. As voice synthesis technology advances, understanding these cognitive boundaries becomes imperative for ethical AI development, system design, and protection of vulnerable populations in voice-mediated interaction.

\begin{acks}
We would like to thank: 
P. Naseck and L. Blanchard for technical advice 
S. Ahmed and N. Whitmore for advice on the study design
A. Schon for help with experiment setup 
and the Study participants for their time

\end{acks}

\newpage
\clearpage
\appendix
\section{Mini-IPIP Questionnaire Items}\label{miniipip}
The Mini-IPIP questionnaire is a short 20-item version of the IPIP-NEO inventory, which assesses the Big Five personality traits:

\textbf{Extraversion}
\begin{itemize}
  \item I am the life of the party.
  \item I don’t talk a lot.
  \item I talk to a lot of different people at parties.
  \item I keep in the background.
\end{itemize}

\textbf{Agreeableness}
\begin{itemize}
  \item I sympathize with others’ feelings.
  \item I am not interested in other people’s problems.
  \item I have a soft heart.
  \item I am not really interested in others.
\end{itemize}

\textbf{Conscientiousness}
\begin{itemize}
  \item I get chores done right away.
  \item I often forget to put things back in their proper place.
  \item I like order.
  \item I make a mess of things.
\end{itemize}

\textbf{Neuroticism}
\begin{itemize}
  \item I have frequent mood swings.
  \item I am relaxed most of the time.
  \item I get upset easily.
  \item I seldom feel blue.
\end{itemize}

\textbf{Openness to Experience}
\begin{itemize}
  \item I have a vivid imagination.
  \item I am not interested in abstract ideas.
  \item I have difficulty understanding abstract ideas.
  \item I do not have a good imagination.
\end{itemize}

\section{Exit Interview}\label{exit-interview}
Our exit interview consisted of the following questions:
\begin{enumerate}[label=\textbf{Q\arabic*:}, leftmargin=*, widest=Q1:]
\item Do you like your own voice?
\item How did you feel hearing altered versions of your voice? Were there moments of discomfort, amusement, curiosity, etc.?
\item Were there any emotional tones in the voice that made it feel more or less like “you”?
\item At what point did the altered voice no longer feel like your own? What specific qualities contributed to that change?
\item Did hearing your voice manipulated in this way change how you think about your own vocal identity?
\item What do you think makes your voice “yours”?
\item Would you trust a system that used your synthetic voice to speak on your behalf?
\item Are you comfortable with technology being able to replicate your voice?
\item Did this experience make you think differently about how others perceive you?
\end{enumerate}

\bibliographystyle{ACM-Reference-Format}
\bibliography{other-references}

\end{document}